\documentclass[showpacs,twocolumn,prl,floatfix]{revtex4}
\usepackage{graphicx}
\usepackage{epsfig}
\usepackage{amssymb}

\def\beq{\begin{equation}}
\def\eeq{\end{equation}}
\def\beqa{\begin{eqnarray}}
\def\eeqa{\end{eqnarray}}

\def\e{\epsilon}

\def\half{{\ss 1\over 2}}

\def\D{\Delta}

\def\e{\epsilon}
\def\cH{{\mathcal H}}
\def\cL{{\mathcal L}}
\def\ss{\scriptstyle}

\def\L{\mathrm{L}}
\def\R{\mathrm{R}}

\def\ss{\scriptstyle}
\def\al{\alpha}

\def\si{\sigma}
\def\etal{{\sl et al.}}

\def\nonum{\nonumber \\}

\def\Tc{T_c}
\def\TL{T_L}
\def\TR{T_R}
\def\TRc{T_{R,c}}
\def\nonum{ \nonumber \\}

\begin{document}

\title{Maintaining the local temperature below the critical value in thermally out of equilibrium superconducting wires}
\author{Y. Dubi and M. Di Ventra}
\affiliation{Department of Physics, University of California San Diego, La Jolla, California
92093-0319, USA} 
\begin{abstract}
A generalized theory of open quantum systems combined with
mean-field theory is used to study a superconducting wire in
contact with thermal baths at different temperatures. It is shown
that, depending on the temperature of the colder bath, the
temperature of the hotter bath can greatly exceed the equilibrium
critical temperature, and still the local temperature in the wire is maintained below the critical temperature and hence the wire remains in the superconducting state. The effects of contact areas and disorder are studied. Finally, an experimental setup is suggested to test our
predictions.
\end{abstract}
\maketitle

Ever since the discovery of superconductivity, fabricating a superconducting (SC) wire that
conducts electricity without dissipation at room temperature has been a major goal of modern
condensed matter physics. However, since the discovery of High-$\Tc$ superconductors
\cite{Bednorz} (which have $\Tc$ as high as $\sim$150K, still far from room temperature and
perhaps close to the upper limit in these materials \cite{Kresin}) there has been little or no
progress in increasing $\Tc$. Recently, several suggestions have been put forward to increase
$\Tc$ by fabricating nanoclusters \cite{Kresin2,Cao} or layering of the SC material
\cite{Yuli,Berg,Okamoto}.

\begin{figure}[h]
\vskip 0.5truecm
\includegraphics[width=8.5truecm]{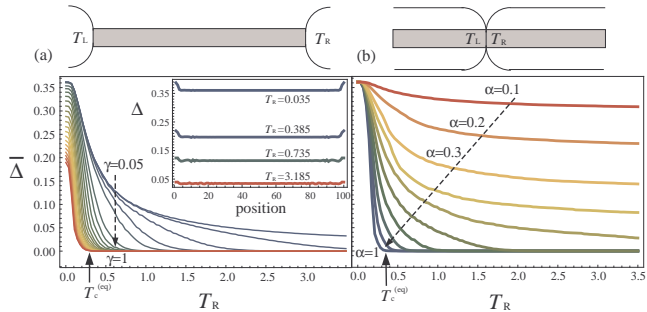}
\caption{(color online) (a) Main panel: $\bar{\D}$ as a function of $\TR$ for different values of
$\gamma$. At small values of $\gamma$, $\TR$ can greatly exceed the equilibrium critical
temperature (indicated by a solid arrow) without destroying superconductivity in the wire. Inset: position dependence of the order parameter for different values of
$\TR$ (at $\gamma=0.05$), demonstrating that it is uniform along the sample. (b) Same as in (a)
but for a constant $\gamma=0.083$ and different values of $\al$ (describing the
contact area of the right heat bath with the sample). Upper panel: the geometries considered in (a) and (b). }\label{fig1}
\end{figure}

 In this paper we consider a different
route which may allow a substantial increase in $\Tc$ for small SC
wires, based on local cooling. To this aim we study a SC wire in
contact with two different heat baths, held at different
temperatures. We consider at first a wire in contact with two heat
baths at its edges, held at different temperatures $\TL$ (left bath)
and $\TR$ (right bath), with $\TR> \Tc^{eq} > \TL$, where $\Tc^{eq}$
is the equilibrium critical temperature (upper panel of
Fig.~\ref{fig1}(a)). We find that depending on the value of $\TL$,
the critical value $\TRc$ (defined as the maximum value of the
temperature $\TR$ of the hot bath at which the wire is still SC) can
be much larger than $\Tc^{eq}$, which implies that the local temperature in the wire is maintained below $\Tc^{eq}$ although the average of $\TL$ and $\TR$ exceeds it. We then study the effect of different
couplings to the baths and of disorder on the above result. We find
the dependence of $\TRc$ on the coupling to the different baths, and
demonstrate that, in agreement with Anderson's theorem
\cite{Anderson}, weak disorder does not change $\TRc$  by much.
Strong disorder leads to the breakdown of the SC state near the
right (hot) bath, and to a spatial dependence of the SC order
parameter.

 Normal metallic wires in contact with two heat baths at the edges were recently studied in detail, in the context of heat flow
in such systems \cite{Pershin,Dubi,Dubi2}. One of the main
conclusions of Ref. \cite{Dubi} was that in a system which is not
diffusive, one cannot define a non-equilibrium temperature as the
average of the temperatures of the different baths. Rather, it is
the energy distribution function (DF) of the two baths which are
averaged (a similar observation was verified experimentally for
short wires, when interactions are relatively unimportant
\cite{Pothier}, a situation which is also likely to hold for the
quasi-particles in SC wires). This observation will allow us to
provide an analytic expression for the critical temperature which
shows excellent fit with our numerical calculations, and provides a
direct prediction which may be tested experimentally.

 The method we use is a generalization of the Bogoliubov-De Gennes (BdG) mean-field theory \cite{DeGennes} to non-equilibrium.
  The starting point is the tight-binding BdG Hamiltonian on a square lattice (with lattice constant $a$=1), $ \cH_{BdG}=\sum_{i,\si}(\e_i-\mu)c^\dagger_{i\si}c_{i \si}-t \sum_{\langle i,j \rangle, \si}
c^\dagger_{i\si}c_{j \si}+ \sum_i \left( \D_i c^\dagger_{i \uparrow} c^\dagger_{i
\downarrow}+h.c. \right)~~, $ where $c^\dagger_{i,\si}$ creates an
electron in the $i$-th lattice site with spin $\si$, $t$ is the hopping integral ($t$=1 serves as
the energy scale hereafter), $\e_i$ are random on-site energies drawn from a uniform distribution
$U[-W/2,W/2]$ (hence $W$ is the strength of disorder, with $W=0$ representing a clean system),
$\mu$ is the chemical potential and $\D_i$ is the SC order parameter in the $i$-th lattice site.
The order parameter is to be determined self-consistently on every site via $\D_i=-U \langle
c_{i\downarrow} c_{i\uparrow} \rangle$, where $U>0$ is the effective electron-electron attractive
interaction and $\langle \cdot \rangle$ stands for a statistical average. In this paper we treat
only s-wave superconductors, but the formalism can easily be extended to account for other kinds
of symmetry.

Since the BdG Hamiltonian is quadratic, it can be exactly diagonalized to describe the
quasi-particle excitations $\gamma_{n \si}$. To diagonalize it, one performs a Bogoliubov
transformation \cite{DeGennes} for the electron operators, $c^\dagger_{i \si}=\sum_n \left(
u_n(i)\gamma^\dagger_{n \si}+\si v^*_n(i)\gamma_{n \bar{\si}} \right) $, where $u_n(i)$ and
$v_n(i)$ are quasi-particle and quasi-hole wave-functions, respectively. Requiring that this
transformation diagonalizes the Hamiltonian yields a set of eigenvalue equations for the
quasiparticle (QP) wave functions \cite{DeGennes}, \beq \left(
  \begin{array}{cc}
    \hat{\xi} & \hat{\D} \\
    \hat{\D}^* & -\hat{\xi}^* \\
  \end{array}
\right)
\left(
  \begin{array}{c}
    u_n(i) \\
    v_n(i) \\
  \end{array}
\right) =E_n\left(
  \begin{array}{c}
    u_n(i) \\
    v_n(i) \\
  \end{array}
\right) ~~,\label{BdGequation} \eeq where $\hat{\xi} u_n(i)=-t \sum_{\langle i,j, \rangle} u_n(j)+(\e_i-\mu+U n_i/2) u_n(i)$
(here $n_i$ is the local electron density), and $\hat{\D}u_n(i)=\D_i u_n(i)$. From Eq. (\ref{BdGequation}) one obtains the QP
wave functions $u_n(i)$ and $v_n(i)$ and the energies $E_n$. In equilibrium, this procedure results in a closed self-consistent
set of equations for the local SC order parameter and density, which were recently
used to study, e.g., effects of disorder and magnetic fields in two dimensional superconductors \cite{Ghosal,Dubi3,Dubi4}.

Since the QP excitations are non-interacting Fermions, they can be
treated by the formalism of Refs.~\cite{Pershin,Dubi,Dubi2}, which
is aimed at studying such particles out of equilibrium. To
generalize the method of Refs.~\cite{Pershin,Dubi,Dubi2} to the QP
excitations, we define a single-particle density matrix $\hat \rho$,
with matrix elements $\rho_{nn'}=\langle
\gamma^\dagger_{n\uparrow}\gamma_{n' \uparrow} \rangle$ (note that
the particle-hole symmetry allows one to treat only the up-spin
excitations). The master equation for $\hat \rho$ is of the Lindblad
form \cite{Lindblad} (setting $\hbar=k_B=1$ hereafter) $
\dot{\rho}=-i[\cH,\rho]+\cL_\L[\rho]+\cL_R[\rho] ~,%\label{LindbladEquation} ~~,\eeq
$ where $\cH$ is the diagonal matrix
of energies $E_n$ and \beq
\cL_{(\L,\R)}[\rho]=\sum_{nn'}\left(-\half
\left\{V^{(\L,\R)\dagger}_{nn'}V^{(\L,\R)}_{nn'},\rho
\right\}+V^{(\L,\R)}_{nn'} \rho V^{(\L,\R)\dagger}_{nn'}\right)
\label{LindbladOperators}\eeq describe environment-induced inelastic
transitions between different single-particle states. The
V-operators in Eq.~(\ref{LindbladOperators}) take on a local form
\cite{Dubi} \beqa V^{(\L,\R)}_{nn'}&=&\sqrt{\Gamma^{(\L,\R)}_{nn'}
f_D^{(\L,\R)}(E_n)} \gamma^\dagger_{n\uparrow}\gamma_{n'\uparrow}
\nonum \Gamma^{(\L,\R)}_{nn'} &=& \left| \Gamma_0 \sum_{r_i \in
S_{\L,\R}} ( u_n(r_i)u^*_{n'}(r_i)+v_n(r_i)v^*_{n'}(r_i)) \right|
~~\label{Voperators} \eeqa where $S_{\L,\R}$ are the contact area of
the left (right) heat bath with the sample,
$f_D^{(\L,\R)}(E_n)=1/\left(1+\exp(E_n/T_{\L,\R}) \right)$ is the
Fermi distribution and $\Gamma_0$ is some constant scattering rate.
(We take $\Gamma_0$ =0.1, changing $\Gamma_0$ does not alter the results presented above.)
Note that the $V$-operators operate on the QPs and not on the
electron operators, since the Fermi function is defined for the
occupation of the QPs (to put it differently, the part of the
electrons which is in the superfluid phase does not feel scattering
from the baths, only the QPs do).

Once the V-operators are evaluated, the master equation is solved in
the asymptotic time limit (i.e., $\dot{\rho}=0$) and a solution for
$\rho$ is obtained. From this solution, the local density and order
parameter are evaluated via the self-consistency condition, which
reads \beqa \D_i &=& U\sum_n u_n(i)v^*_n(i) (1-2\rho_{nn}) \nonum
n_i&=&2\sum_n\left(|u_n(i)|^2\rho_{nn}+|v_n(i)|^2(1-\rho_{nn})
\right)~~. \label{SelfConsistency}\eeqa With $\D_i$ and $n_i$
determined, the whole procedure (i.e., the solution of the BdG
equations, the evaluation of the V-operators and the solution of the
master equation) is repeated until  $\D_i$ and $n_i$ no longer
change (within the numerical tolerance of
 $<10^{-5}$). We point that this treatment is of a mean-field type,
and as such neglects statistical fluctuations in the Hamiltonian
\cite{Dagosta} or phase-slips \cite{phaseslips}. As we will show
below, in the ideal case one can define an effective temperature of
the wire, and hence in such a case one can follow the usual
treatment of phase-slips in SC wires, but using the effective
temperature as input. Phase fluctuations are unlikely to change the
local effective temperature, since these are excitations that do not
carry heat (as opposed to QP excitations). In the disordered case,
such a treatment cannot work as the temperature becomes
space-dependent and one needs a different formalism to treat
phase-fluctuations in the presence of a temperature gradient. Such a
theory is beyond the scope of the present work and will be the
subject of future studies.

We begin by presenting the averaged order parameter
$\bar{\D}=\frac{1}{N} \sum_i \D_i$, where $N$ is the total number of
lattice sites. We consider the geometry shown in the side panel of
Fig.~\ref{fig1}(a), where the SC wire (gray area) is connected to
the thermal baths only at its edges. The numerical parameters are as
follows. The wire dimensions are $100 \times 10, U=2, W=0$ (clean
system) and the density is held at $n=0.875$ (i.e., $n$ electrons
per site on average, and the chemical potential is chosen
self-consistently to maintain this filling). We define $\TL$ by its
ratio with the critical temperature at equilibrium $\Tc^{eq}$,
$\TL/\Tc^{eq}=\gamma<1$.

If Fig.~\ref{fig1}(a), $\bar{\D}$ is plotted as a function of $\TR$ for different values of
$\gamma$, $\gamma=0.05$, (top curve), $0.1,..., 1$ (bottom curve). For low values of $\gamma$ ,
$\TRc$ can greatly exceed $\Tc^{eq}$ (marked by a solid arrow). In the inset of Fig.~\ref{fig1}(a)
we plot the local order parameter as a function of position along the wire (averaged over the
transverse direction), at $\gamma=0.05$ for different values of $\TR=0.035, 0.385, 0.735$ and
$3.185$ (in units of $t$). We find that although there are two different temperatures at the
edges, the order parameter is practically uniform along the wire, in agreement with the results of
Refs.~\cite{Dubi,Dubi2}.

 In Fig.~\ref{fig1}(b) we study a somewhat different (and perhaps more realistic) situation, in which the heat baths are in contact with the wire not only at the edges but over some area (see side panel of Fig.~\ref{fig1}(b)). We define the parameter $\al$ ($1-\al)$ to be the ratio between the
contact area of the right (left) heat bath and the area of the whole wire,
such that $\al=1$ stands for a system in full contact only with the
right heat bath. In Fig.~\ref{fig1}(b), $\bar{\D}$ is plotted as a
function of $\TR$ (at $\gamma=0.083$) for different values of $\al$,
$\al= 0.1$, (top curve), $0.2,...,1$ (bottom curve). The $\al =1$
curve is the equilibrium curve (with $\Tc^{eq}$ marked by an arrow).
As seen, for different values of $\al$, $\TRc$ may again exceed
$\Tc^{eq}$. Also in this case we found that both $\D$ and the local
temperature (calculated for a similar one-dimensional geometry, with
the method of Refs.~\cite{Dubi,Dubi2}) are uniform in space (not
shown). This emphasizes the fact that the temperature is defined not
by the local baths, but rather by the (inelastic) scattering between
states, which in the clean case span the entire system.

If one could simply define a local temperature which gradually shifts from $\TL$ to $\TR$, then
one would expect that $\TR$ could not exceed $\Tc^{eq}$ and that the order parameter would not be
uniform in space. In order to explain our findings, we recall that one of the main results of Refs.
\cite{Dubi,Dubi2} is that in the ballistic limit the temperature is uniform, and a non-equilibrium
DF develops, which is the average of the two DFs of the left and right baths. In the case
represented in Fig.~\ref{fig1}(b), we find that a weighted average between the DFs of the left and
right baths develops, the weight being $\al$. This result, along with the observation that $\D_i$
is uniform in space, allows us to find an analytical expression for the effective $\Tc$ as
follows. In the equilibrium theory of superconductivity \cite{DeGennes}, the critical temperature
$\Tc^{eq}$ is determined by the gap equation $ \frac{1}{N_0
U}=\frac{1}{2}\int^{\omega_D/\Tc^{eq}}_0 \frac{1}{x} (1-2 f(x)) dx~~, $%\label{Gap Equation}~~,\eeq
where $N_0$ is the density of states at the Fermi energy, $\omega_D$ is the Debye frequency and
$f(x)$ is the DF. From the above discussion, in the non-equilibrium case we have $f(x)=\al
f^{(\R)}_D(x)+(1-\al) f^{(\L)}_D(x)$. The resulting equation for $\TRc$ then reads
 \beqa \frac{1}{N_0
U}&=&\frac{\al}{2}\int^{\omega_D/\TRc}_0 \frac{1}{x} (1-2 f(x)) dx+\nonum &
&+\frac{1-\al}{2}\int^{\omega_D/\TL}_0 \frac{1}{x} (1-2 f(x)) dx
 \label{Non Eq Gap Equation}~~.\eeqa
These integrals may be evaluated exactly, and with
$\TL/\Tc^{eq}=\gamma$ we find \beq \frac{\TRc}{\Tc^{eq}}=\gamma^{1-1/ \al}\label{Tc}~~.\eeq

\begin{figure}[h!]
\vskip 0.5truecm
\includegraphics[width=8.5truecm]{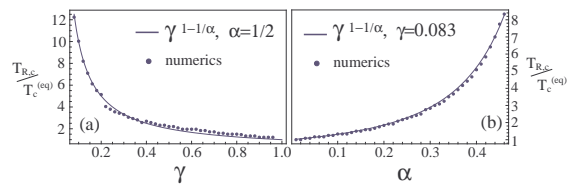}
\caption{(a) The ratio $\TRc/\Tc^{eq}$ as a function of $\gamma$, taken from the data of
Fig.~\ref{fig1}(a). The points correspond to the numerical data and the solid line is
Eq.~(\ref{Tc}) with $\al=0.5$. (b) $\TRc/\Tc^{eq}$  as a function of $\al$, taken from the data of
Fig.~\ref{fig1}(b). The points correspond to the numerical data and the solid line is
Eq.~(\ref{Tc}) with $\gamma=0.083$.}\label{fig2}
\end{figure}

 In Fig.~\ref{fig2} we plot $\TRc/\Tc^{eq}$ as a function of $\gamma$ (Fig.~\ref{fig2}(a)) and of $\al$
(Fig.~\ref{fig2}(b)), taken from the data of Fig.~\ref{fig1}(a) and(b), respectively. The solid
line corresponds to Eq.~(\ref{Tc}) for the two cases, with the corresponding parameters taken from
the numerical calculation. The agreement between Eq.~(\ref{Tc}) and the numerical
results confirms that indeed the DF is a weighted average of the DFs of the two baths. One can now
use Eq.~(\ref{Tc}) to estimate the effective $\TRc$ of real materials. As a practical use, it is
advantageous to raise $\TRc$ above the freezing point of liquid Nitrogen, $\sim 77K$. For example,
consider a desired working temperature of $\sim 80$K, and local refrigerators which cool down to
$40K$, deposited on a SC wire with $\Tc^{eq} \sim 60$K. A cover of 40\% refrigerators would
increase $\TRc$ to $=78$K.

Perhaps a more intriguing possibility is the enhancement of $\TRc$ to
room temperature. For a wire made of the newly-found Iron compound
\cite{Kamihara,Takahashi,Grant} ($\Tc^{eq} \sim 50$K) heated (or
cooled) at the edges ($\al =0.5$), a temperature $\TL=7.5$K would
drive $\TRc$ above room temperature. For nano-scale wires made of a
high-$T_c$ material, the fabrication of which was recently
demonstrated \cite{Xu}, taking $\Tc^{eq} \sim 80$K, local cooling of
$\TL=20$K and coverage of $\al=0.5$ would drive $\TRc$ above room
temperature. Of course, other effects (e.g. phonon scattering, phase fluctuations etc.) might become very important in such high value of $\TR$ and inhibit the zero-resistance state.

\begin{figure}[h!]
\vskip 0.5truecm
\includegraphics[width=6.5truecm]{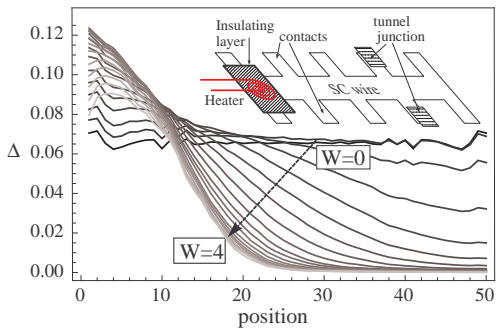}
\caption{Position dependence of the order parameter $\D$, calculated with increasing values of
disorder, from the clean case $W=0$ (dark line) to strongly disordered case $W=4$ (light gray
line), averaged over $500$ realizations of disorder.
For this example $\TL=0.02$, $\Tc^{eq}=0.2$ and $\TR=1.5$.
Inset: Suggested experimental setup. An insulating layer is deposited on top of a SC
wire with etched contacts, and a heater coil is placed on top of it, to generate local heating.
The resistance is then monitored along the other contacts (which may be either 4-terminal contacts
or tunnel junctions).}\label{fig3}
\end{figure}

Next we turn to study the effect of disorder. In Ref.~\cite{Dubi2}
it was shown that the form of the non-equilibrium DF is robust
against disorder, but that the local temperature profile changes
from a constant-temperature to a position-dependent profile. In
Fig.~\ref{fig3} we plot the order parameter $\D_i$ as a function of
position along the wire for $\TR=1.5$ (which is above $\Tc^{eq} \sim
0.2$ in this example), $\TL=0.02$ and dimensions $50 \times 10$, for
different values of disorder, $W=0$ (dark curve), $0.2,...,
4$ (bright curve). The order parameter is averaged over $500$
realizations of disorder. For a clean system $\D$ is uniform, and
assumes a position-dependence with increasing disorder. For small
values of disorder it is finite everywhere in the sample, in
agreement with Anderson's theorem \cite{Anderson}. For large values
of disorder it vanishes near the right edge, and increases at the
left edge. This is due to the fact that a local temperature ensues
which varies from the left to the right temperature. In fact, if one compares Fig.~\ref{fig3} to the local temperature calculated for similar (normal) systems (Fig.~1 of Ref.~\cite{Dubi2}) one finds that $\D$ obeys a simple BCS-like law, with the temperature replaced by the local temperature.

We point out
that in this case the local temperature near the left edge (which
corresponds to $\TL$) is lower than the effective temperature of the
clean sample, and hence the rise in the value of $\D$ with
increasing disorder near the left edge. The distance from the right
edge at which $\D$ vanishes indicates the "thermal length" where the
local temperature is close to $\TR$ \cite{Dubi2}. We also note that
already for relatively small values of disorder ($W \approx 0.5$ in
our case) the localization length is smaller than the system size,
which means that the onset of a vanishing gap in the system
(occurring at $W \approx 1.4$)  does not directly correspond to the
onset of localization \cite{Dubi2,Dubi5}.

In order to test our predictions, we suggest the experimental setup
shown in the inset of Fig.~\ref{fig3}. It consists of a SC wire with
etched contacts (either regular 4-terminal contacts, or made as
tunnel junction, aimed at measuring the local density of states at
that location). On top of one of the contacts an insulating layer is
deposited, and on top of it a heater coil is set. The temperature
below the heater can be calibrated by measuring the resistance
between the contacts beneath the heater when the wire is in the
normal state. Then the whole device is cooled down, and the
resistance through the other contacts is measured as a function of
the current that passes through the heater (i.e., the local
temperature beneath it). In a uniform system, the resistances of all
the contacts should vanish if the system is SC, and have finite
values once the local temperature beneath the heater rises above
$\TRc$. If the system is disordered, the different contacts should
exhibit a gradual transition to a normal state.

The length-scale that determines the onset of a temperature
gradient is in this case the QP mean-free path (or diffusion
length)\cite{Dubi5}. While it can be estimated for low $\Tc$ metals,
for high-$\Tc$ materials it is unknown (although it is suspected to
be small, based on their poor conduction in the normal state). By
controlling the distances between the contacts our proposed
experiment can thus serve to determine this length, by relating it
to the length-scale at which a temperature gradient develops.
%\begin{figure}[h!]
%\vskip 0.5truecm
%\includegraphics[width=6truecm]{fig4.eps}
%\caption{Suggested experimental setup. An insulating layer is deposited on top of a SC wire with etched contacts, and a heater
%coil is placed on top of it, to generate local heating. The resistance is then monitored along the other contacts (which may be
%either 4-terminal contacts or tunnel junctions).}\label{fig4}
%\end{figure}

In this work we have neglected the effect of phonon scattering, and assumed the effective electron-electron interaction is not appreciably changed by  temperature, supported by the fact that the Debye temperature is much larger than $\Tc$. However, considering our geometry, electron-phonon (e-ph) interaction effects may play a significant role as $\TR$ reaches the Debye temperature. This is probably more important in disordered wires (for clean wires we expect that, on equal grounds, phonons will also acquire a uniform temperature). The e-ph interaction may induce inelastic electron transitions, which will reduce the inelastic mean-free path. Since the temperature profile is sensitive to the inelastic mean-free path \cite{Dubi5}, this effect may be seen in our suggested experiment.

To conclude, we point out that in recent years there have been
tremendous advances in fabricating micro-refrigerators, based on the
thermo-electric Peltier effect, and which can locally cool down
their environment substantially \cite{Shakouri,Chowdhury}.  Since
the efficiency of thermo-electric materials is likely to increase in
the future \cite{Majumdar}, one can conceive a device composed of a
(relatively) high-$\Tc$ material, on top of which are embedded a series of micro-refrigerators, covering an area of the material and cooling it enough for it to operate at a temperature which exceeds its $\Tc$, a possibility that may allow for integrating superconducting wires as circuit elements in various devices.

We are grateful to A. Sharoni for valuable discussions. This research was funded by DOE under grant DE-FG02-05ER46204.

\end{document}